\documentclass[twocolumn,amsmath,amssymb,aps,prc,floatfix]{revtex4-1}
\usepackage[english]{babel}
\usepackage{amsfonts}
\usepackage{graphicx}
\usepackage{color}
\usepackage{subfigure}
\usepackage{hyperref}
\usepackage{setspace}    %Zeilenabstand auf 1.5 Zeilen ändern
\usepackage{multirow}
\begin{document}

\title{Baryon-antibaryon annihilation and reproduction in relativistic heavy-ion collisions}
\author{E. Seifert}
\affiliation{Institut f\"{u}r Theoretische Physik, Universit\"{a}t Gie\ss en, Germany}

\author{W. Cassing}
\affiliation{Institut f\"{u}r Theoretische Physik, Universit\"{a}t Gie\ss en, Germany}
    \date{\today}
    \begin{abstract}
The quark rearrangement model for baryon-antibaryon annihilation and
reproduction ($B {\bar B} \leftrightarrow 3 M$) - incorporated in
the Parton-Hadron-String Dynamics (PHSD) transport approach - is
extended to the strangeness sector. A derivation of the transition
probabilities for the three-body processes is presented and a
strangeness suppression factor for the invariant matrix element
squared is introduced to account for the higher mass of the strange
quark compared to the light up and down quarks. In simulations of
the baryon-antibaryon annihilation and reformation in a box with
periodic boundary conditions we demonstrate that our numerical
implementation fulfills detailed balance on a channel-by-channel
basis for more than 2000 individual $2 \leftrightarrow 3$ channels.
Furthermore, we study central Pb+Pb collisions within PHSD from
11.7$A$\,GeV to 158$A$\,GeV and investigate the impact of the
additionally implemented reaction channels in the strangeness
sector. We find that the new reaction channels have a visible impact
essentially only on the rapidity spectra of antibaryons. The spectra
with the additional channels in the strangeness sector are closer to
the experimental data than without for all antihyperons. Due to the
chemical redistribution between baryons/antibaryons  and mesons we
find a slightly larger production of antiprotons thus moderately
overestimating the available experimental data. We additionally
address the question if the antibaryon spectra (with strangeness)
from central heavy-ion reactions at these energies provide further
information on the issue of chiral symmetry restoration and
deconfinement. However, by comparing transport results with/without
partonic phase as well as including/excluding effects from chiral
symmetry restoration we find no convincing signals in the strange
antibaryon sector for either transition due to the strong
final-state interactions.

        \par
        PACS: 24.10.-i; 24.10.Cn; 24.10.Jv; 25.75.-q; 14.65.-q
    \end{abstract}
    \maketitle

\section{Introduction}

Lattice Quantum-Chromo-Dynamics (lQCD) calculations suggest that at
vanishing baryon chemical potential ($\mu_B$=0) there is a crossover
phase transition from hadronic to partonic degrees of freedom
\cite{lQCD,lqcd0,LQCDx,Peter,Lat1,Lat2} for the deconfinement phase
transition as well as for the restoration of chiral symmetry. This
leaves the open question whether or not a first-order phase
transition might occur at finite baryon chemical potential implying
a critical endpoint in the QCD phase diagram \cite{CBMbook}. Since
lattice calculations so far suffer from the fermion-sign problem,
model-independent information on the QCD phase diagram can presently
only be obtained from experimental data. It is thus expected that a
thorough study of this issue with relativistic heavy-ion collisions
(HIC) of different system sizes and at various bombarding energies
will provide further information. However, the problem here is the
model dependence in the interpretation of the measured particle
yields and their relation to the properties of the fireball created
in the collision \cite{Stock}.

Among the many observables suggested the strangeness enhancement was
already proposed in the 80's of the past century
\cite{Rafelski:1982pu} as a probe of the Quark-Gluon-Plasma (QGP).
Particularly hyperons and antihyperons should provide an ideal
sample for QGP fireballs since in the initial colliding nuclei no
net strange quarks are present and a major part of the produced
$s\bar s$ pairs should be produced by gluon fusion processes in the
QGP \cite{Koch:1986ud}. However, due to a partial restoration of
chiral symmetry close to the hadron-parton transition the $s\bar s$
production threshold is lowered in a dense hadronic medium and
strangeness enhancement might also signal chiral symmetry
restoration rather than deconfinement as suggested in Refs.
\cite{Soff,Cass16,NICA,AlesPaper}. We note that quark confinement and
chiral symmetry breaking are not intimately connected at finite
$\mu_B$ \cite{Redlich}. The multi-strange baryons and antibaryons
are expected to be more sensitive to the QGP than single-strange
baryons or mesons since multi-strange baryons, and particularly
multi-strange antibaryons, are suppressed by high hadronic energy
thresholds as well as by long timescales for multi-step processes in
a purely hadronic phase \cite{Blume,Soff,stock2}. This, however, holds only
for two-body production channels whereas three-body channels (e.g.
by three vector mesons) do not suffer from severe energy thresholds.
Accordingly, the reaction dynamics for baryon-antibaryon ($B {\bar
B}$) annihilation and recreation in the hadronic phase have to be
under control before solid conclusions can be drawn on the boundary
in the QCD phase diagram or on freeze-out conditions in relativistic
heavy-ion reactions. Furthermore, all strangeness exchange channels
in the hadronic phase have to be taken into account as pointed out
in Refs. \cite{Ko1,Ko2}.

A first step in this direction has been taken in Ref.
\cite{Cassing:2001ds} where the three-body fusion of nonstrange
pseudoscalar and vector mesons to $B {\bar B}$ pairs has been
incorporated in the Hadron-String Dynamics (HSD) transport approach
\cite{HSD} that preferentially describes the hadronic phase. Here
the matrix element squared has been extracted from the experimental
data on $p {\bar p}$ annihilation and the three-body meson channels
have been determined on the basis of detailed balance. It was found
that in central collisions of heavy nuclei the annihilation of
antinucleons is almost compensated by the inverse recreation
channels. If this holds true also in the strangeness sector is
presently unknown. Furthermore, the former HSD calculations did not
incorporate a deconfinement phase transition to the QGP nor effects from chiral
symmetry restoration and thus did not
allow to draw any conclusions on the phase boundary of QCD.

On the other hand the HSD transport approach has been further
extended in the last 15 years  a) to the formation of an initial partonic phase with quark and gluon
quasiparticle properties that are fitted to lattice QCD results in thermodynamic
equilibrium, b) to a dynamical hadronization scheme on the basis of covariant
transition rates, c) to incorporate further hadronic reactions
in the strangeness sector with full baryon-antibaryon symmetry
and d) to employ essential aspects of chiral symmetry
restoration in the hadronic phase \cite{AlesPaper}. Whereas the latter
developments are important for the lower Super Proton Synchrotron (SPS) energy regime to
account for the strangeness enhancement seen experimentally in heavy-ion collisions,
the formation of a partonic phase is mandatory to understand the physics at higher
SPS, Relativistic Heavy-Ion Collider (RHIC) and Large Hadron Collider (LHC) energies.
Since multistrange baryons and antibaryons at top  SPS
energies no longer stem from string fragmentation (as in HSD \cite{Cassing:2001ds})
but preferentially from hadronization at energy densities around 0.5 GeV/fm$^3$ the
issue of three-meson fusion reactions for the formation of baryon-antibaryon ($B {\bar B}$)
pairs and the annihilation of  $B {\bar B}$ pairs to multiple mesons has to be reexamined.

In this work we will present, furthermore, the extension of the quark
rearrangement model (QRM) for baryon-antibaryon annihilation and
recreation to the strangeness/antistrangeness sector (briefly denoted by SU(3)).
We will show the impact of these additional reaction channels
for heavy-ion collisions using the Parton-Hadron-String Dynamics
(PHSD) transport approach to simulate central Pb+Pb collisions in
the bombarding energy regime from 11.7$A$\,GeV to 158$A$\,GeV. The PHSD
\cite{PRC08,PHSD,Bratkovskaya:2011wp}, which incorporates in
addition to HSD a transition to the partonic phase as well as
dynamical hadronization, reproduces many observables for p+p, p+A
and A+A collisions ranging from SPS up to Large Hadron Collider
(LHC) energies \cite{AlesPaper,review,Linnyk:2015tha,Konchakovski:2012yg}.
Since PHSD is found to also
well describe the spectra of strange mesons and baryons from
heavy-ion collisions from 2$A$\,GeV up to RHIC/LHC energies when
incorporating aspects of chiral symmetry restoration in the hadronic
phase \cite{AlesPaper}, its performance in the strange antibaryon
sector will be tested using the extended QRM and also lead to
predictions for rare multi-strange baryons and antibaryons in the
lower energy regime where experimental data are scarce or lacking at
all.

This work is organized as follows: In Sec. \ref{sec:PHSD} we
recapitulate shortly the ingredients of PHSD while in Sec.
\ref{sec:QuarkRearrangement} we briefly recall and motivate the
quark rearrangement model  for baryon-antibaryon annihilation and
recreation ($B\bar B\leftrightarrow 3M$). We extend the QRM to the
strangeness sector and introduce a strange quark suppression factor
for the transition matrix element squared in the strangeness sector.
After deriving the transition probabilities on the basis of detailed
balance, we present in Sec. \ref{sec:box} the validity of our
numerical implementation for detailed balance in case of $B\bar
B\leftrightarrow 3M$ reactions including the strangeness sector
within simulations in a finite box with periodic boundary
conditions. In Sec. \ref{sec:PHSDsimulation} we present results for
antibaryons and multi-strange baryons from PHSD simulations for
central Pb+Pb collisions in the  SPS energy regime and study the
impact of chiral symmetry restoration and deconfinement. We will
compare simulations using the baryon-antibaryon annihilation and formation
with and without the strangeness sector with each other
and to available experimental data for rapidity and transverse mass
spectra. Furthermore, we compare the PHSD results for central Pb+Pb
reactions at  40$A$\,GeV with those from the Ultra-relativistic
Quantum Molecular Dynamics model (UrQMD) \cite{Petersen:2008kb} and
the three-fluid dynamics model (3FD) using a 2-phase equation of
state \cite{Ivanov:2013yqa}. We conclude our study with a summary in
Sec. \ref{sec:summary} while more technical details are described in
the appendices.

\section{The PHSD transport approach}\label{sec:PHSD}
The PHSD is a microscopic
covariant transport approach for strongly interacting systems which
is based on  Kadanoff-Baym equations
\cite{Kadanoff-Baym,Schwinger:1960qe,Schwinger-Keldysh2,Botermans:1990qi}
for the Green's functions in phase-space representation in first
order gradient expansion \cite{Cassing:1999wx,Cassing:1999mh}. Due
to its basis on the Kadanoff-Baym equations it can describe systems
in and out-of equilibrium and goes beyond the quasiparticle
approximation by incorporating dynamical spectral functions for the
partons. It is capable of describing the equilibration process of
systems which are far out-of equilibrium to the correct equilibrium
state \cite{Juchem:2003bi}. The PHSD incorporates a partonic as
well as a hadronic phase to describe all stages of a relativistic
heavy-ion collision with transitions from strings to dynamical
partons as well as dynamical hadronization. In the hadronic phase
high-energy resonance decays are described by multi-particle string
decays.
%The degrees-of-freedom differ in the two phases as it has
%partonic degrees of freedom, i.e. quarks, antiquarks and gluons, in
%the partonic phase and baryons, mesons and strings in the hadronic
%phase. With this recipe
PHSD is capable of simulating the full time evolution of a
relativistic heavy-ion collision - from impinging nuclei in their
'groundstates' to the final hadronic particles - ranging from
SchwerIonen-Synchroton (SIS), Alternating Gradient Synchrotron (AGS)
over Facility for Antiproton and Ion Research (FAIR)/
Nuclotron-based Ion Collider fAcility (NICA) up to Relativistic
Heavy-Ion Collider (RHIC) and Large Hadron Collider (LHC) energies
and is able to reproduce a large number of observables in these
energy regimes for p+p, p+A and A+A reactions
\cite{AlesPaper,review}.

The properties of the off-shell partonic degrees-of-freedom are
determined by the Dynamical-Quasi-Particle-Model (DQPM)
\cite{Peshier:2004bv,Peshier:2004ya} which provides the masses, widths and spectral
functions of the dynamical gluons and quarks/antiquarks \cite{PeshierPRL,Crev}. The
(essentially three) parameters of the DQPM are chosen to reproduce
the lQCD equation-of-state at vanishing baryon chemical potential.
It has been shown that using PHSD in a box with periodic boundary
conditions it reproduces the lQCD results for transport coefficients
such as the shear and bulk viscosity as well as the electric
conductivity for the partonic phase \cite{Vitaly,Cass13,Steinert}.

In the PHSD simulation of a nucleus-nucleus collision the primary
hard nucleon-nucleon scatterings  produce strings which are
color-singlet states described by the FRITIOF Lund model
\cite{NilssonAlmqvist:1986rx}. As the strings decay they produce
"pre-hadrons" that have a formation time of $\tau_f\approx 0.8\,$fm
while "leading hadrons", which originate from the string ends, may
interact instantly without  formation time but with reduced
cross-sections in line with the constituent quark model \cite{HSD}.

If the local energy density $\epsilon$ is above the critical value
of $\epsilon_c \approx 0.5$\,GeV/fm$^3$, as provided by lQCD
calculations, the unformed hadrons dissolve into dynamical quarks
with properties defined by the DQPM at given energy density. In the
partonic phase these partons propagate in the scalar self-generated
mean-field potential and scatter with each other with cross sections
extracted from the dynamical widths of partons. The expanding system
then leads to a decreasing local energy density until it is close to
or below the critical value $\epsilon_c$. At this point the partonic
degrees-of-freedom  hadronize to colorless off-shell mesons and
baryons by the fusion of massive quark-antiquark pairs or the fusion
of three quarks (antiquarks) conserving energy, three-momentum and
quantum numbers in each event \cite{PRC08}. In the hadronic phase -
as found in the corona and at late reaction times - the hadrons
interact with each other in elastic and inelastic collisions with
cross sections taken from experimental data or evaluated within
effective hadronic Lagrangian models. The detailed balance relation
for each reaction channel is incorporated and ensures the correct
backward reaction rates. In particular the strangeness exchange
reactions are included in meson-baryon/antibaryon, baryon-baryon
and antibaryon-antibaryon collisions
following Refs. \cite{Song1,Song2,Song3}. Furthermore, retarded
electromagnetic fields as generated by the electric charge currents
(from charged hadrons and quarks) are incorporated \cite{Toneev17}.

\begin{figure}[t]
\centering
\includegraphics[width= 0.48\textwidth]{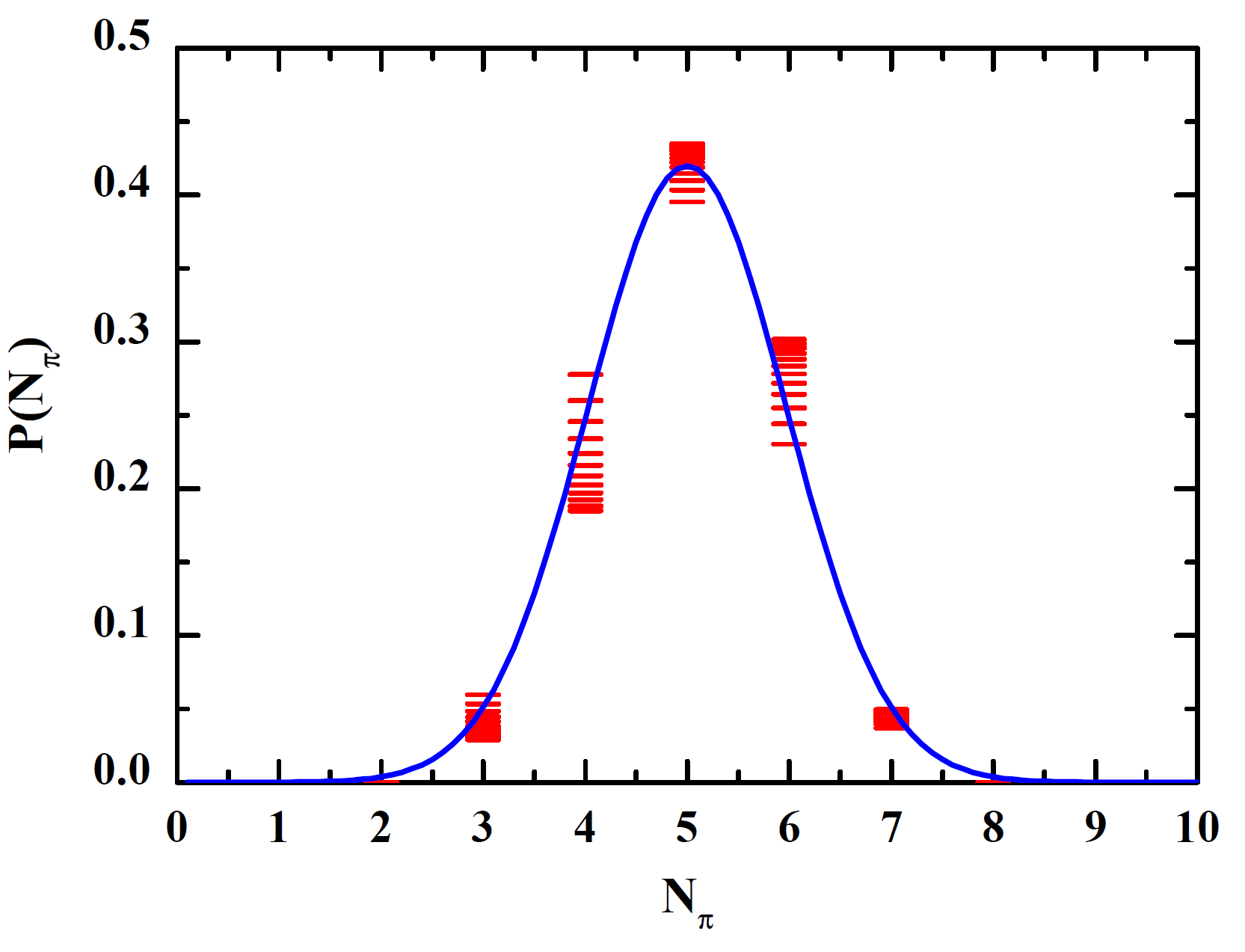}
\caption{(Color online) Distribution in the final number of pions
$P(N_\pi)$ for $p\bar p$ annihilation at invariant energies
$2.3\,\mathrm{GeV}\leq\sqrt s \leq 4\,\mathrm{GeV}$ from the QRP
(short lines). The solid line is a gaussian parametrization fitted
to the experimental data. The figure is taken from Ref.
\cite{Cassing:2001ds}.}\label{fig:piondistribution}
\end{figure}

\section{Quark rearrangement model for $B+\bar B$ production and
annihilation}\label{sec:QuarkRearrangement}
In this section we present the quark rearrangement model and the
most relevant equations for the two- and three-body scattering
rates. An extensive description for the invariant reaction rates for
general particle number changing processes as well as the motivation
for the quark rearrangement model is given in Ref.
\cite{Cassing:2001ds}.

\subsection{Concept}
As discussed in Ref. \cite{Cassing:2001ds} one experimentally finds
a dominant annihilation of $p\bar p$ into 5 pions at invariant
energies $2.3\,\mathrm{GeV}\leq\sqrt s \leq 4\,\mathrm{GeV}$, see
Fig. \ref{fig:piondistribution}.
\begin{figure}[b]
{\centering
\includegraphics[width= 0.48\textwidth]{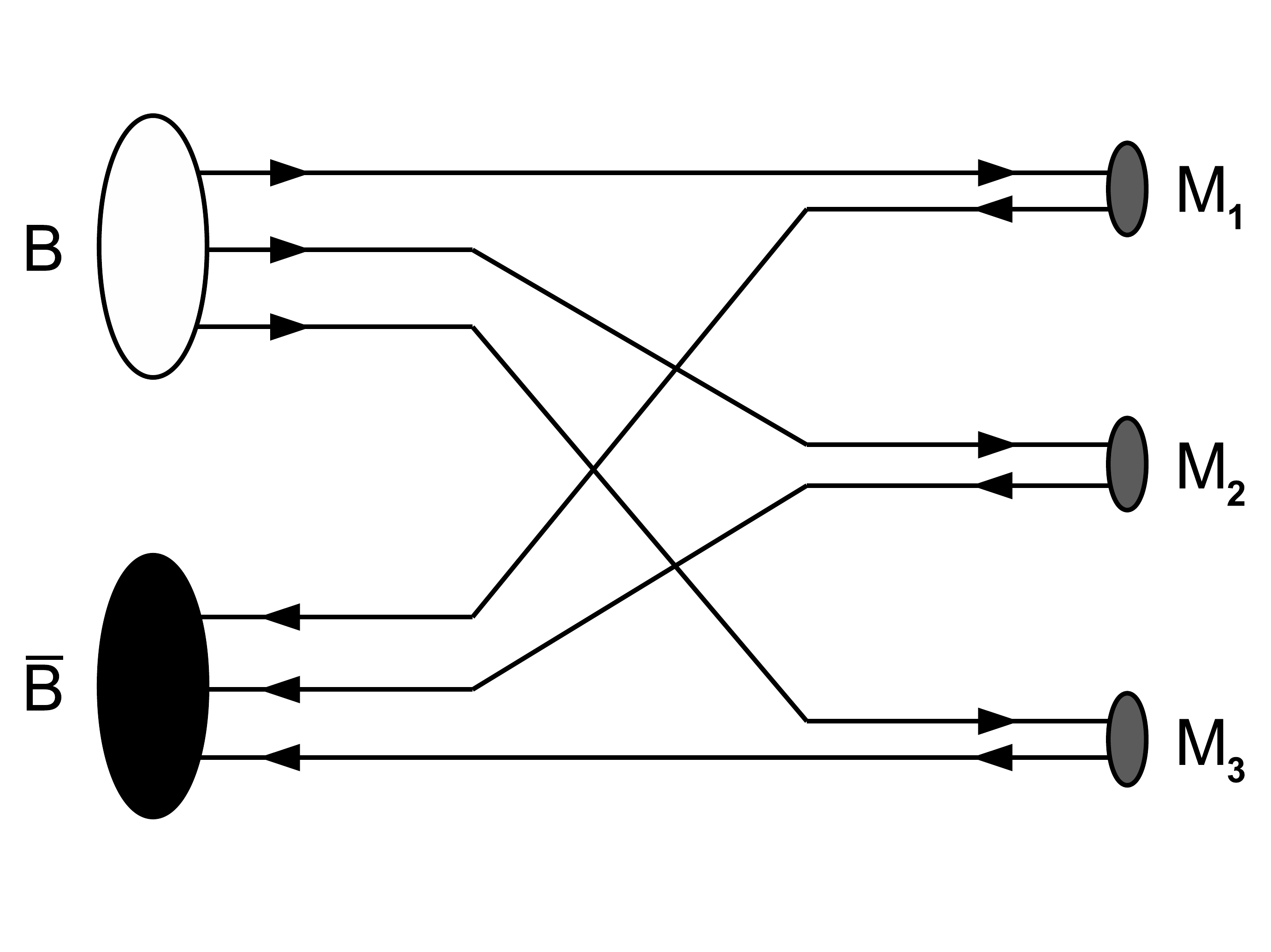}}
\caption{Illustration of the quark rearrangement model for a general
baryon antibaryon pair $B\bar B$ annihilating into three mesons $M$
and vice versa. Here the meson $M_i$ may be any of the $0^-$ or $1^-$
nonets. }\label{fig:rearrangement}
\end{figure}
The final number of 5 pions may be interpreted as an initial
annihilation into $\pi\rho\rho$ with the $\rho$ mesons decaying
subsequently into two pions each. The channel  $\pi\pi\rho$ then
leads to 4 final pions, the channel  $\pi\omega\rho$ to 6 final
pions, the channel  $\rho\omega\rho$ to 7 final pions etc.
Accordingly, the baryon-antibaryon annihilation in the first step is
a two-to-three reaction with a conserved number of quarks and
antiquarks. This is the basic assumption of the quark rearrangement
model which is also illustrated in Fig. \ref{fig:rearrangement}. The
annihilation reaction $p\bar p\rightarrow \pi\rho\rho$ is the
dominant process in $p\bar p$ annihilation for invariant masses
below 4\,GeV, typical for the hadronic phase of a heavy-ion
collision. By allowing the mesons $M_i$ to be any member  of the $0^-$
or $1^-$ nonets one can describe an arbitrary $B \bar B$
annihilation and recreation by rearranging the quark and antiquark
content.  An implementation of baryon-antibaryon annihilation in
such a manner misses the annihilation into one or two mesons,
however,  higher numbers of final mesons are implemented through the
subsequent decay channels. This approach gives a realistic
description for $p {\bar p}$ annihilation and we assume that for
other baryon-antibaryon pairs than $p\bar p$ a similar annihilation
pattern holds. Since there are no measurements of annihilation cross
sections other than $p\bar n$ and $p\bar p$ this is our best guess
which might be falsified by experiment.

\subsection{Covariant transition rates}
The quark rearrangement model only contains reactions of the kind
$2\leftrightarrow 3$. The detailed balance based Lorentz invariant
on-shell collision rate for the reaction $B\bar B\rightarrow 3M$ in
a volume element of size d$V$ and time-step size d$t$ is written as
\cite{Cassing:2001ds}:
\begin{multline}
\frac{\mathrm{d}N_\mathrm{coll}[B\bar B \rightarrow 3~\mathrm{mesons}]}{\mathrm{d}t\,\mathrm{d}V}=\\
\sum_c\sum_{c'}\frac{1}{(2\pi)^6}\int\frac{\mathrm{d}^3p_1}{2E_1}\frac{\mathrm{d}^3p_2}{2E_2}~W_{2,3}(\sqrt{s})\\
\times   R_3(p_1+p_2;c) N_\mathrm{fin}^c f_1(x,p_1)f_2(x,p_2).\label{eq:lorentzinvariant23}
\end{multline}
In (\ref{eq:lorentzinvariant23}) $c'$ denotes all $B\bar B$ pairs
with the properties $c'=(m_1^{c'},m_2^{c'};\nu^{c'})$; $c$ are all
the possible meson channels with $c=(m_3^c,m_4^c,m_5^c;\lambda^c)$,
with $m$ being the masses of the respective particles, and $\nu$ and
$\lambda$ the quantum numbers signifying the channel (charge,
parity, spin and strangeness). We assume that the transition matrix
element squared $W_{2,3}$ does not significantly depend on the
outgoing momenta and just on the invariant mass of the reaction,
which holds approximately  true for $p\bar p$ as we will see later.
A formulation based on the matrix element will ensure detailed
balance. The on-shell $n$-body phase-space integral is defined by
\begin{align}
\begin{split}
 R_n(P;m_1,\ldots,m_n)=\left(\frac{1}{(2\pi)^3}\right)^n  &\int\prod_{k=1}^n \frac{\mathrm{d}^3p_k}{2E_k}~(2\pi)^4
 \\ &\hspace{-0.2cm}\times \delta^4\left(P-\sum_{j=1}^n p_j\right)\label{eq:n-phasespace}
\end{split}\end{align}
and in case of a constant transition matrix element dominates the
interaction rate of the system. The factor $N_\mathrm{fin}^c$ is the
multiplicity of the meson triple $c$ and results from the summation
over the spin $s$ and possible isospin projections $F_\mathrm{iso}$
compatible with charge conservation of the meson channel $c$:
\begin{align}
N_\mathrm{fin}^c=(2s_3+1)(2s_4+1)(2s_5+1)\frac{F_\mathrm{iso}}{N_\mathrm{id}!}.
\end{align}
The division by $N_\mathrm{id}!$, with $N_\mathrm{id}$ denoting the
number of identical mesons,  ensures that each charge configuration
is only considered once for a given meson triple. The functions $f$
are the distribution functions of the $B\bar B$ pair in momentum and
coordinate space. When looking at a specific $B\bar B$ pair one has
to make sure that only meson channels are considered which conserve
charge, energy and parity. The probability of this specific $B\bar
B$ pair $c'$ to annihilate into any of these possible meson channels
$c$ is related to the total annihilation cross section of the $B\bar
B$-pair $\sigma_\mathrm{ann}^{c'}$ \cite{kajantie}:
\begin{gather}
\begin{split}
\frac{P^{c'}_\mathrm{tot}\mathrm{d}V}{\mathrm{d}t}=&\frac{1}{4E_1E_2}
\sum_c W_{2,3}(\sqrt s)R_3(p_1 + p_2;c)N_\mathrm{fin}^c\\=&v_\mathrm{rel}\sigma^{c'}_\mathrm{ann}(\sqrt s),\end{split}\label{eq:P23}\\
v_\mathrm{rel}=\frac{\sqrt{\lambda(s,m_1^2,m_2^2)}}{2E_1E_2} ~;~ \lambda(a,b,c)=(a-b-c)^2-4bc , \label{eq:lambdadef}
\end{gather}
where $dV$ and $dt$ are taken finite.
The probability for a specific final state  $\tilde P^{c'\rightarrow
c}$ in case of an annihilation is then given by the available
phase space and the multiplicity of all possible meson channels $c$:
\begin{gather}
\tilde P^{c'\rightarrow c} = N_3(c,c',\sqrt s)R_3(p_1+p_2;c)N_\mathrm{fin}^c, \\
\mathrm{with} \quad N_3^{-1}(c,c',\sqrt s)=\sum_c R_3(p_1+p_2;c)N_\mathrm{fin}^c. \label{eq:N3}
\end{gather}
In a similar manner one finds for the probability of a specific
meson channel $c$ fusing together and forming a specific $B\bar B$
pair $c'$,
\begin{align}
\begin{split}
\frac{P^{c\rightarrow c'} \mathrm{d}V^2}{\mathrm{d}t} =
&\frac{1}{4E_3E_4E_5}\sigma_\mathrm{ann}^{c'}(\sqrt s)N_3(c,c',\sqrt
s)\\ &\times\frac{\lambda(s,m_1^2,m_2^2)}{8\pi s}N_B^{c'},
\end{split}\label{eq:P32}
\end{align}
with $N_B^{c'}=(2s_1+1)(2s_2+1)$ denoting the multiplicity of the
$B\bar B$ pair. A more detailed derivation of these formulae is
given in Appendix \ref{sec:mesonfusion}.

\subsection{Annihilation cross sections}
\begin{figure}[b]
{\centering
\includegraphics[width= 0.48\textwidth]{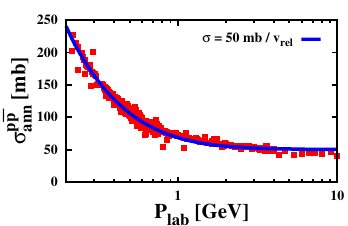}}
\caption{(Color online) $p\bar p$ annihilation cross section as a function
of momentum in the laboratory $P_{lab}$. The data points are
taken from \cite{Agashe:2014kda} and the solid line is a fit by the
function 50 mb/$v_\mathrm{rel}$ with $v_\mathrm{rel}$ denoting the
relative velocity in the laboratory
system (\ref{eq:lambdadef}).}\label{fig:ppbarcrosssection}
\end{figure}

For the calculation of actual collision probabilities, Eq.
(\ref{eq:P23}) and (\ref{eq:P32}), we are still missing the cross
sections. As already mentioned above we assume the cross sections to
depend only on the invariant energy, not the outgoing momenta. This
assumption is approximately fulfilled for $p\bar p$ and $p\bar n$ annihilation,
see Fig. \ref{fig:ppbarcrosssection}. Other channels have not been
measured so far. Since there are no experimental data available we
assume a similar behavior for different spin combinations like
$p\bar \Delta$.
%Additionally, we incorporate a cut at a maximum of
%80\,mb due to in-medium screening effects. This limits the
%interaction length to about 1.6\,fm.

In this work we investigate in
particular the strangeness sector. We model the cross sections of
particles with strangeness by
\begin{align}
\sigma^{c'}_\mathrm{ann}(\sqrt s)=\sigma^{p\bar p}_\mathrm{ann}\lambda^{\varsigma + \bar\varsigma}, \label{eq:crosssection}
\end{align}
where $\varsigma$ and $\bar\varsigma$ are the number of strange and
antistrange quarks in the $B\bar B$ pair $c'$ and $\lambda\in [0,1]
$ is a factor suppressing the transition matrix element for
particles taking part in the quark rearrangement model and
effectively suppressing the cross section. This parametrization is
motivated by  PYTHIA \cite{PYTHIA} simulations where one sees a
similar suppression for particles with strangeness compared to
non-strange particles at the same energy above threshold. In the
final implementation in PHSD the suppression factor has the value
$\lambda$=0.5 which is in rough agreement with the PYTHIA
simulations embedded in PHSD. We choose a dependence on not just the
net strangeness $S$ but the sum of strange and antistrange quarks
$\bar\varsigma + \varsigma$ due to their higher mass  and a
subsequent suppression of the rearrangement. The implementation with
the strangeness $|S|=|\bar\varsigma - \varsigma|$ instead of
$\bar\varsigma + \varsigma$ has no practical influence on the final
results in case of relativistic heavy-ion collisions (cf. Appendix
\ref{sec:strangeness}).

\section{Simulations in a finite box}\label{sec:box}
\begin{figure}[tb]
{\centering
\includegraphics[width= 0.48\textwidth]{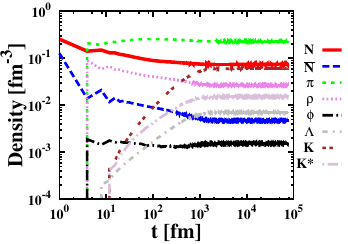}}
\caption{(Color online) Particle densities of a $p\bar p$
initialized system as a function of time. The particle species
correspond to the following lines: the red solid line corresponds to
nucleons, the blue dashed line to  antinucleons, the green
short-dashed line to pions, the violet dotted line to  $\rho$
mesons, the black dashed-dotted line to  $\phi$ mesons, the grey
dashed-doubly-dotted line to  $\Lambda$s, the brown doubly-dashed
line to  kaons and the beige short-dashed-dotted line to the vector
kaons $K^*$. The different charge states of the particles have been
summed over and $K$ denotes the sum of $K^+,~K^-,~K^0$ and $\bar
K^0$.}\label{fig:densities}
\end{figure}
This section addresses the implementation of the $2 \leftrightarrow
3$ reactions  formulated above in PHSD and checks the consistency of
the numerical implementation. We use transport simulations in a box
with periodic boundary conditions to investigate the behavior of the
quark rearrangement model in equilibrium. We recall that in
equilibrium - according to detailed balance - the reaction rate for
$B\bar B\rightarrow 3M$ should be the same as for $3M\rightarrow
B\bar B$. Furthermore, for a consistent implementation detailed
balance should not only be fulfilled for the sum of all reaction
channels but on a channel by channel basis. In the box simulations
only the $B\bar B\leftrightarrow 3M$ reactions are considered now
and all particles are taken as stable such that no decays occur. The
particles incorporated are the $0^-,1^-$ meson nonets, the $1/2^+$
baryon octet and the $3/2^+$ baryon decuplet. Additionally, we
consider N(1440) and N(1535) baryonic resonances. Furthermore, we
take into account the strangeness content of $\eta$ and $\phi$ with
50\% and 83.1\% $s\bar s$ content, respectively. With this the
number of possible mass channels amounts to more than 2000. Hence,
an initialization with every possible channel is not feasible.
Therefore, we look at systems which are initialized by a single type
of baryon and  antibaryon adding up to 100 systems for the
consistency check. The box simulations have the following initial
conditions:
\begin{itemize}
\item Box volume lies around 18000\,fm$^3$ with periodic boundary conditions
\item All simulations have the same energy density $\epsilon = 0.4 $\,GeV/fm$^3$ with 10\% of the energy distributed to kinetic energy
\item The ratio between baryons and antibaryons is set to 2:1 and the net baryon density amounts to  $\rho_B \approx 0.2$\,fm$^{-3}$
\item The initial momentum distribution is of Boltzmann-shape
\item For the box simulations a suppression of channels including strangeness is neglected.
\end{itemize}
\begin{figure}[tb]
{\centering
\includegraphics[width= 0.48\textwidth]{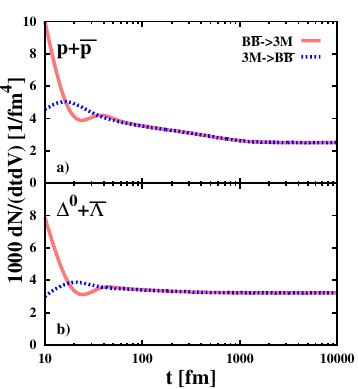}}
\caption{(Color online) Total reaction rate (per volume $dV$)
as a function of time for two different
initializations. The solid (slightly transparent) red lines
correspond to the baryon-antibaryon annihilation and the red dotted
lines to the formation. The systems are initialized in a) with only
$p+\bar p$ and in b) with only $\Delta^0+\bar
\Lambda$.}\label{fig:rate-t}
\end{figure}

\begin{figure}[b]
{\centering
\includegraphics[width= 0.48\textwidth]{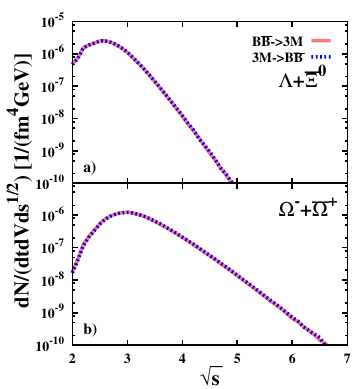}}
\caption{(Color online) Total reaction rate as a function of the invariant mass $\sqrt s$ in
equilibrium for five different initializations. The solid (slightly transparent)
red line corresponds to the baryon-antibaryon annihilation and the red dotted
line to the formation. The systems are initialized in a) with only
$\Lambda+\bar\Xi^0$ and in b) with only $\Omega^- +\bar\Omega^+$.}\label{fig:rate-s}
\end{figure}

Before the actual calculations the three-body phase-space integrals $R_3$
have been calculated and fitted by proper functionals to save enormous
CPU time. In detail, the three-body phase-space integrals $R_3$, depending on invariant energy
and three masses $m_i$,
\begin{multline}
R_3(\sqrt s;m_1,m_2,m_3)=\\ \int\limits_{(m_1+m_2)^2}^{(\sqrt s-m_3)^2}
\frac{\mathrm{d}M_2^2}{2\pi} R_2(\sqrt s;m_3,M_2)R_2(M_2;m_1,m_2),
\end{multline}
with $R_2$ defined in Eq. (\ref{eq:R2}) are fitted by
\begin{align}
R_3(t,m_1,m_2,m_3)=a_1 t^{a_2}\left( 1-\frac{1}{a_3 t+1+a_4}\right),
\end{align}
with $t=\sqrt s -m_1-m_2-m_3$ and $a_i > 0$. The fit parameters $a_i$
have been evaluated  for each combination of meson masses $m_1,m_2,m_3$
and stored on file. For further  details  on the phase-space integrals
we refer the reader to  appendix \ref{sec:phasespace}.

We recall that the fusion of three mesons can not be described in a
Lorentz invariant way by geometrical collision criteria between the
particles due to the three inertial systems. To find a solution we
employ the in-cell method introduced by Lang et al.
\cite{in-cell-method} and adopted in Ref.
\cite{Cassing:2001ds}. This method is also employed
for $2 \leftrightarrow 3$ reactions in partonic cascade calculations
\cite{Greiner}. The in-cell method can be used for any number of
colliding particles since there is no problem with time ordering due
to the locality of the formulation. In the in-cell method space-time
is divided into four dimensional cells with widths $\Delta x,\Delta
y,\Delta z,\Delta t$ and only particles inside the same cell may
collide with each other. One calculates the reaction probabilities
of each particle with every other one inside the same cell. The
actual collision and the final state is chosen via Monte-Carlo. The
possible final states and multiplicities in Eqs. (\ref{eq:P23}) and
(\ref{eq:P32}) are precalculated to save computational time during
the transport simulation. The cell size and the time step $\Delta t$
are optimized for the problem under investigation such that the
total probability of a transition in a local cell does not exceed
unity but is also not too small. For the actual calculations shown below
we use $dt$ = 4 fm/c and $dV$=40 fm$^3$ which ensures that the transition probabilities
are always below unity.

\begin{table*}[t]
\caption{Deviation from detailed balance $\delta$ (\ref{eq:deviation-detailed-balance}) for selected
systems and as the average over all 100 investigated systems
$\langle \delta\rangle$.}\label{tab:deviation-detailed-balance}
\centering
\begin{tabular}{c||l|r||l|r||l|r||c}
\multicolumn{1}{c||}{\multirow{2}{*}{rank}} & \multicolumn{2}{c||}{$p+\bar p$} & \multicolumn{2}{c||}{$\Delta^0+\bar \Lambda$} & \multicolumn{2}{c||}{$\Lambda + \bar\Xi^0$} & \multicolumn{1}{c}{\multirow{2}{*}{\small $\langle\delta\rangle$ [\%]}} \\ \cline{2-7}

\multicolumn{1}{c||}{} & channel & {\small $\delta$ [\%]} & channel & {\small $\delta$ [\%]} & channel & {\small $\delta$ [\%]} & \multicolumn{1}{c}{} \\ \hline\hline
% 1 15 48

1 & {\small $N \bar N\leftrightarrow \pi\pi\rho$} & 0.17 & {\small $N\bar\Xi\leftrightarrow \pi K K^*$} & 1.45 & {\small $N\bar N \leftrightarrow \pi\pi\rho$} & 0.13 & 1.24\\
2 & {\small $N\bar N \leftrightarrow \pi\rho\rho $} & 3.06 & {\small $N\bar\Omega \leftrightarrow K K^* K^* $} & 3.59 & {\small $N\bar\Delta \leftrightarrow\pi\rho\rho $} & 1.70 & 1.82\\
3 & {\small $N\bar \Delta\leftrightarrow\pi\pi\rho $} & 1.58 & {\small $\Delta \bar \Xi \leftrightarrow \pi K K^* $} & 1.32 & {\small $N\bar\Delta\leftrightarrow\pi\pi\rho $} & 2.04 & 1.70\\
4 & {\small $N\bar\Delta\leftrightarrow\pi\rho\rho $} & 0.84 & {\small $\Delta\bar\Xi\leftrightarrow K K^*\rho $} & 0.64 & {\small $N\bar N\leftrightarrow\pi\rho\rho $} & 3.31 & 1.54\\
5 & {\small $\Delta\bar N \leftrightarrow\pi\pi\rho $} & 2.43 & {\small $\Delta\bar\Omega\leftrightarrow K K^* K^* $} & 1.08 & {\small $\Delta\bar N\leftrightarrow \pi\rho\rho $} & 1.33 & 1.49\\
6 & {\small $\Delta \bar N \leftrightarrow\pi\rho\rho $} & 0.73 & {\small $N\bar\Sigma\leftrightarrow \pi K^*\rho $} & 3.58 & {\small $\Delta \bar N \leftrightarrow\pi\pi\rho $} & 2.71 & 1.97\\
7 & {\small $N\bar N\leftrightarrow \pi\pi a_1 $} & 6.52 & {\small $\Delta\bar \Sigma\leftrightarrow \pi K^*\rho $} & 2.00 & {\small $\Delta\bar\Delta\leftrightarrow\pi\pi\rho $} & 2.69 & 2.04\\
8 & {\small $N\bar N\leftrightarrow\pi\pi\pi $} & 5.10 & {\small $N\bar N \leftrightarrow \pi\pi\rho $} & 0.23 & {\small $N\bar \Sigma\leftrightarrow \pi K^*\rho $} & 2.04 & 2.03\\
9 & {\small $N\bar\Sigma\leftrightarrow \pi K \rho $} & 0.31 & {\small $N\bar\Sigma\leftrightarrow \pi K \rho $} & 0.42 & {\small $\Delta\bar\Delta\leftrightarrow\pi\pi\rho $} & 2.12 & 2.11\\
10 & {\small $N\bar\Sigma\leftrightarrow \pi K^*\rho $} & 0.96 & {\small $N\bar\Omega\leftrightarrow K K K $} & 0.35 & {\small $N\bar\Sigma\leftrightarrow \pi K \rho $} & 0.35 & 2.11\\
\end{tabular}
\end{table*}

%\subsection{Results}
We now discuss results for a few selected systems. We present
randomly picked ensembles that cover the qualitative range of
possible systems, i.e. systems consisting of only initial light
quarks, only initial strange/antistrange  quarks as well as a variety of combinations of
light and strange quarks/antiquarks. In Fig. \ref{fig:densities} the time
evolution of the particle densities for a system initialized with
protons and antiprotons is shown to demonstrate the production and
annihilation of different particle species in a system consisting
initially only of protons and antiprotons. After the first timestep
of the simulation a lot of new mesons like pions, $\rho$ and
$\omega$ mesons are formed. At later times also strange mesons and
baryons are formed because of the partial $s\bar s$ content of
$\phi$ and $\eta$. In equilibrium the system has a significant
amount of mesons and baryons with strange and antistrange quarks.
However, the generation of strange quarks even for the meson sector
takes a long time ($\approx$ 60\,fm/c) to produce significantly high
strange particle densities; thus the generation via $\phi$ and
$\eta$ should have negligible influence on actual heavy-ion
collisions since large densities are needed for a significant
contribution from the meson fusion. In a 5\% central Pb+Pb collision
at 158$A$\,GeV the meson fusion dies out at $\sim$ 13\,fm/c which is
insufficient for having a major influence on the strangeness sector,
see Fig. \ref{fig:phsd-rate-t} (discussed in section \ref{sec:PHSDsimulation} below).

We show in Fig. \ref{fig:rate-t} the total reaction rate as a
function of time for two exemplary initializations which were
initialized with $p+\bar p$ and $\Delta^0+\bar \Lambda$,
respectively.
%,~ \Lambda+\bar\Xi^0,~ \Sigma^- + \bar\Omega^+$ and $\Omega^- +\bar\Omega^+$from top to bottom.
Both systems share a similar evolution of the total reaction rate.
All systems reach detailed balance much faster ($\approx$ 40\,fm)
than they reach equilibrium ($\approx$ 1000\,fm).

Detailed balance should also be valid for the total reaction rate as
function of the invariant mass. For this we show in Fig.
\ref{fig:rate-s} the total reaction rate as a function of the
invariant mass $\sqrt s$ in the plateau region of Fig.
\ref{fig:rate-t} which is associated with the equilibrium state.
From Fig. \ref{fig:rate-s} we see that  detailed balance
is also fulfilled for this quantity. Note that the maximum
achievable invariant mass of particles participating in annihaltion or
recreation (in equilibrium) is lower in systems initialized with
lighter baryons than for systems initialized with heavier ones.

The last most crucial check for detailed balance is the fulfilment
on a channel by channel basis. To this end we define the deviation from
detailed balance for each channel by
\begin{align}
\delta = 1-\frac{\frac{\mathrm d N}{\mathrm d t}(B\bar B\rightarrow 3M)}{\frac{\mathrm d N}{\mathrm d t}(3M\rightarrow B\bar B)} .\label{eq:deviation-detailed-balance}
\end{align}
We calculate $\delta$ for each of the more than 2000 channels and look at the channels with
the largest reaction rates in all 100 investigated systems. In Tab.
\ref{tab:deviation-detailed-balance} the 10 most important channels
with the largest reaction rates are shown from highest to lowest for
3 of the  exemplary systems as well as the average for all 100
investigated systems and the average over all channels. The average over
all 100 investigated systems shows that detailed balance is
fulfilled better than 97\% on a channel-by-channel basis for the 100
most dominant channels. This verifies the correct implementation of
the baryon-antibaryon annihilation and recreation within the quark
rearrangement model in the PHSD transport approach. Some channels of a
system may deviate by more than 5\% from detailed balance, however,
this is a relict of too low statistics. We found only few channels
($\approx$ 20 for the 10 most dominant channels) that had a
deviation of up to 9\%. In general these deviations may be neglected
as can be seen in the averaged values and the dominant number of
channels being very close to detailed balance which gives a proof for
the working principle of the implementation presented.

\section{PHSD simulations for heavy-ion collisions}\label{sec:PHSDsimulation}
In this section we show the influence of the additional channels in
the strangeness sector for $B\bar B\leftrightarrow 3M$ reactions on
heavy-ion collisions in the energy regime of 11.7-158$A$\,GeV.
%There
%is a difference of technical nature between the old/ standard
%implementation and the new one including strangeness in the
%properties of the particles considered for $B\bar B \leftrightarrow
%3M$ reactions. The new selection of partners is now in line with the
%rest of PHSD.

\begin{figure}[b]
 {\centering
 \includegraphics[width= 0.48\textwidth]{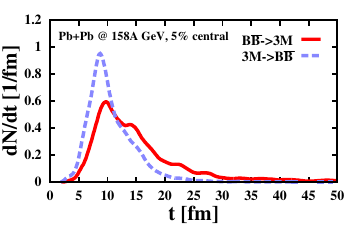}}
 \caption{(Color online) The reaction rate of the $B\bar B\leftrightarrow 3M$ reactions (solid line) as a function of time in
  5\% central Pb+Pb collisions at 158$A$\,GeV in comparison to the total three-meson fusion rate (dashed line).
}\label{fig:phsd-rate-t}
\end{figure}
\begin{figure*}[t]
\includegraphics[width= 0.33\textwidth]{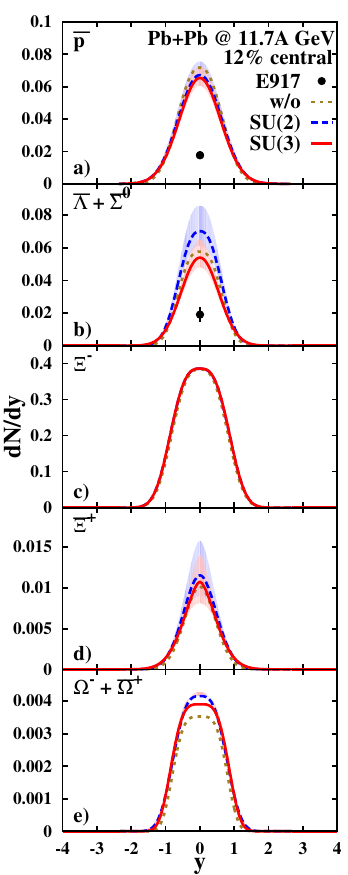}\includegraphics[width= 0.33\textwidth]{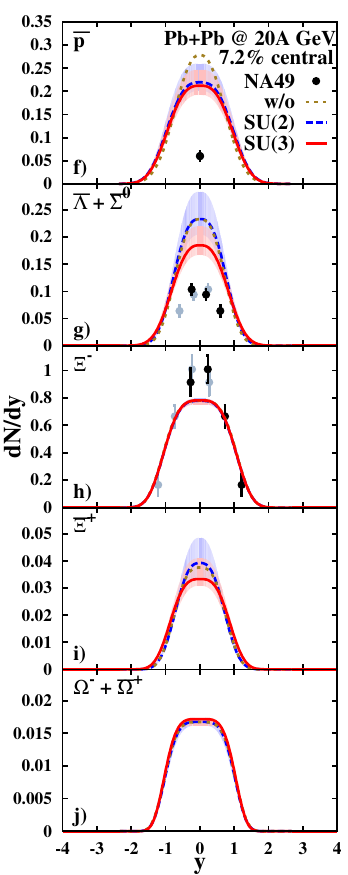}
\includegraphics[width= 0.33\textwidth]{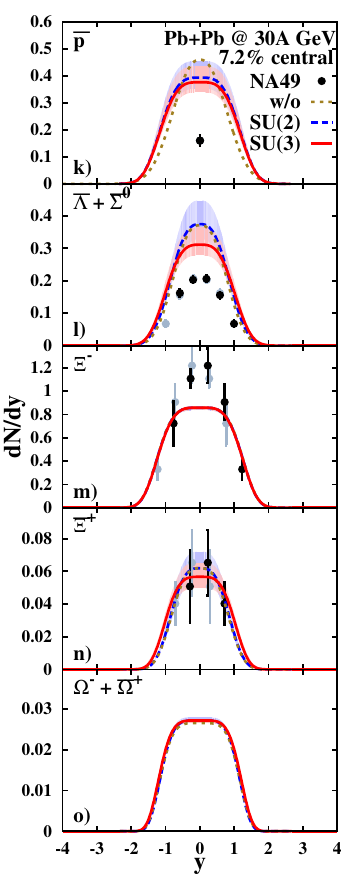}
\caption{(Color online) Rapidity spectra of ${\bar p}, {\bar
\Lambda}+{\bar \Sigma}^0, \Xi^-, {\bar \Xi}^+, \Omega^- + {\bar
\Omega}^+$ in (12\%) 7.2\% central Pb+Pb collisions at 11.7, 20 and
30$A$\,GeV. The solid lines show the results when including all
light and strange quark channels (denoted by SU(3)) while the dashed
lines results from discarding strange or antistrange quarks in the
reaction channels (denoted by SU(2)). The error bands indicate the
systematic uncertainty of the calculations due to a different
ensemble size. The dotted lines show the results with
$B\bar B\leftrightarrow 3M$ reactions switched off.
The data points are taken from Refs.
\cite{Back:2001ai,Anticic:2010mp,Alt:2008qm}.}\label{fig:energyrange1}
\end{figure*}

\begin{figure*}[t]
\includegraphics[width= 0.33\textwidth]{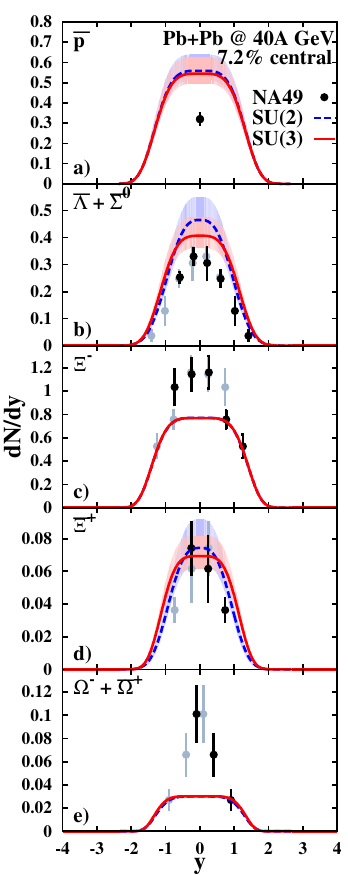}\includegraphics[width= 0.33\textwidth]{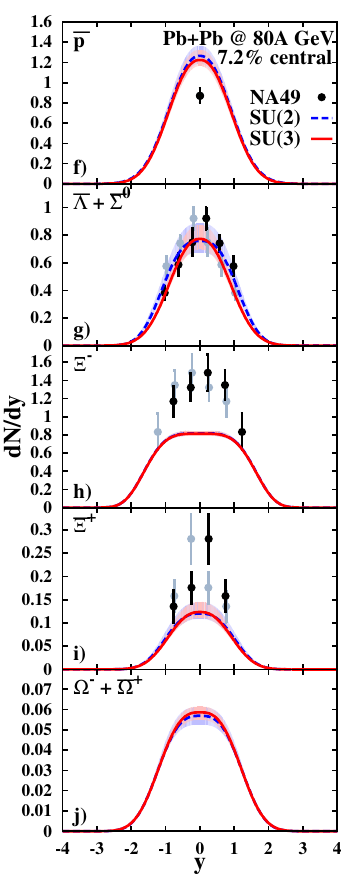}
\includegraphics[width= 0.33\textwidth]{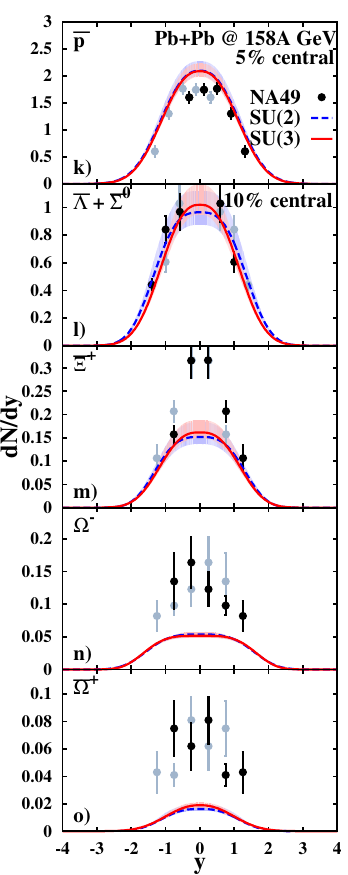}
\caption{(Color online) Rapidity spectra ${\bar p}, {\bar
\Lambda}+{\bar \Sigma}^0, \Xi^-, {\bar \Xi}^+, \Omega^- + {\bar
\Omega}^+$ in central Pb+Pb collisions at 40, 80 and 158$A$\,GeV for
$B\bar B\leftrightarrow 3M$ with only light quarks (dashed lines)
and including strange quarks (solid lines) compared to experimental
measurements. The error bands indicate the systematic uncertainty of
the calculations due to a different ensemble size.
%From top to
%bottom the rapidity spectra of $\bar
%p,\bar\Lambda+\bar\Sigma^0,\Xi^-,\bar\Xi^+$ and
%$\Omega^-+\Omega^+$/$\Omega^-,\bar\Omega^+$ are shown.
The data points are taken from Refs.
\cite{Anticic:2010mp,Alt:2008qm,Alt:2004kq,Alt:2006dk}.}\label{fig:energyrange2}
\end{figure*}

\begin{figure}[t]
{\centering
\includegraphics[width= 0.45\textwidth]{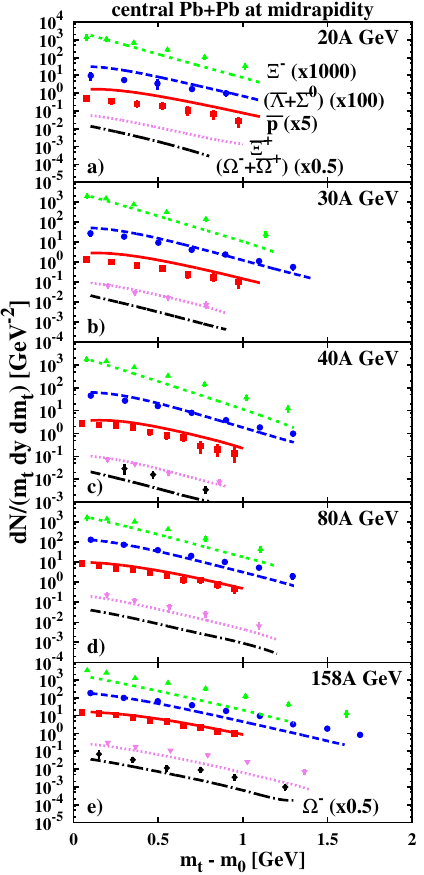}}
\caption{(Color online) Transverse mass spectra for central Pb+Pb
collisions at midrapidity. The centrality selection for the
particles at the different energies is the same as in Figs.
\ref{fig:energyrange1} and \ref{fig:energyrange2}. The particles in
each panel are from top to bottom
$\Xi^-,\bar\Lambda+\bar\Sigma^0,\bar p,\bar\Xi^+$ and
$\Omega^-+\bar\Omega^+$, only at 158$A$ GeV the lowest lying line
corresponds to $\Omega^-$. The data points are taken from Refs.
\cite{Alt:2008qm,Alt:2004kq,Alt:2006dk}.}\label{fig:mtspectrum}
\end{figure}

Before coming to the actual results we compare in
Fig. \ref{fig:phsd-rate-t} the reaction rate for the total
baryon-antibaryon annihilation (solid line) and formation (dashed
line) from PHSD in 5\% central Pb+Pb collisions at 158$A$\,GeV. Whereas the
meson-fusion rate dominates at early times ($<$ 13 fm/c) the
annihilation takes over for larger times during the final expansion
of the system. Although the time integrals of both rates are about
the same there is no appreciable time interval in which both rates
are identical. This indicates a strong nonequilibrium dynamics of
baryon-antibaryon annihilation and reproduction in actual heavy-ion
reactions.

We note that a similar analysis has been performed in the earlier study
in Ref. \cite{Cassing:2001ds} (Fig. 7) on the basis of the HSD transport model
(version 2.3) for the same system, however, without averaging over the ensembles.
The earlier rates differ substantially from the present results from PHSD (version 4.0)
due to the different degrees of freedom in the initial phase of the collision. In
order to quantify the differences we have recalculated the rates within HSD2.3
(from the year 2002) and compared the numbers with those from PHSD4.0, which is
the most recent version including also the effects from chiral symmetry restoration
\cite{AlesPaper} (PHSD3.3) and nonperturbative charm dynamics as well as extended $2
\leftrightarrow 3$ reactions. We
found that both rates (from HSD2.3 and PHSD4.0) differ only slightly for
times $\geq$ 6 fm/c (after contact of Pb+Pb at b=2fm) but the huge rates (from HSD2.3)
at the first few fm/c are essentially missing in PHSD4.0. This is due to the fact
that at the top SPS energy the initial energy conversion goes to interacting partons
in PHSD4.0 and not to strings decaying to hadrons (and partly to $B {\bar B}$ pairs) in HSD2.3.
Thus in PHSD4.0 (at the top SPS energy) there are initially no $B {\bar B}$ pairs that
might annihilate nor mesons that might fuse! Due to the very high hadron densities in HSD2.3
(after string decay) both the annihilation and reproduction rates are very high and about
equal whereas in the hadronic expansion phase the densities are sizeably lower. In this
dilute regime the 3-body channels first dominate and decrease fast in time whereas the 2-body
annihilation reactions still continue for some time. As addressed in the Introduction we thus
expect also differences in the antibaryon rapidity spectra as compared to the early results
from HSD2.3 \cite{Cassing:2001ds}.
However, in both transport calculations -- incorporating the $2 \leftrightarrow 3$ reactions --
the time integrated rates for annihilation and reproduction turn out to be about equal.

The actual PHSD calculations for relativistic nucleus-nucleus
collisions are carried out in the parallel ensemble method, i.e. in
case of the cascade mode a typical number of 100 - 300 ensembles are
propagated in time fully independent from each other. However, the
calculation of net-baryon densities, scalar densities and energy
densities - needed for the full PHSD dynamics - is carried out by
averaging over all ensembles. This results in a crosstalk between
ensembles due to the propagation of particles in the self-generated
mean fields (for partons and baryons/antibaryons) as well as in the
baryon/antibaryon formation in the hadronization. A systematic study
of all particle spectra in rapidity and transverse mass shows that
the results for mesons and baryons well scale with the number of
ensembles whereas the antibaryon sector shows small variations with
the number of ensembles. This scaling violation is essentially due
to the numerical approximations that have to be presently introduced
in order to keep the huge number of reaction channels manageable.
This introduces a systematic error in our calculations for the
antibaryon sector which is accounted for by hatched bands in the
following figures. The solid or dashed lines correspond to the
standard ensemble number of 150 used as default in PHSD calculations
in the energy range of interest.

\subsection{Rapidity and transverse mass spectra}
We now discuss the influence of the $B {\bar B} \leftrightarrow 3 M$
reactions on observables measured in actual experiments from 11.7 -
158$A$\,GeV. We  first focus on rapidity spectra and mention that
the $B\bar B\leftrightarrow 3M$ reactions have practically no
influence on baryon and meson spectra \cite{AlesPaper} and, hence,
we only show the results for the relevant antibaryons and $\Xi^-$ to
demonstrate that the influence on baryons is barely visible. For
results on meson and baryon spectra we refer the reader to the
review \cite{review} and Ref. \cite{AlesPaper}. As mentioned above
the full, dashed and dotted lines show the results for 150 ensembles; the
blue and red hatched areas result when employing different ensemble
numbers in a wide range.

We first focus on the influence of the newly incorporated
strangeness sector. In the following, we  compare the implementation
with only light quark channels (SU(2)) with the new one including
also the strangeness sector (SU(3)). The rapidity spectra of ${\bar
p}, {\bar \Lambda}+{\bar \Sigma}^0, \Xi^-, {\bar \Xi}^+, \Omega^- +
{\bar \Omega}^+$ for central Pb+Pb collisions from 11.7 to
158$A$\,GeV are shown in Figs. \ref{fig:energyrange1}  and
\ref{fig:energyrange2}. The rapidity spectra of the anti-hyperons
are overall closer to the experimental data when taking into account
the strangeness sector for the $B\bar B\leftrightarrow 3M$
reactions. However, the antiproton spectra are faintly influenced by
the incorporated sector and describe the data only moderately well. In
general the investigations suggest that the $B \bar B\leftrightarrow
3M$ reactions have the largest impact at energies below 80$A$\,GeV.
This result shows that the consideration of the strange quarks helps
improving the description of a heavy-ion collision in the framework
of PHSD. For particles like $\bar \Xi^+,\Omega^-$ and $\bar\Omega^-$
at lower energies, where currently no experimental data are
available, our results should be taken as predictions.

In Fig. \ref{fig:energyrange1} we, furthermore, show results
from calculations neglecting the $B\bar B\leftrightarrow 3M$ reactions.
We find that the rapidity distribution for $\bar p$
has a higher peak and is narrower compared to calculations with
$B\bar B\leftrightarrow 3M$, while the total number of antiprotons
is about the same. The results for the antihyperons - starting from 20$A$\,GeV -
lie on top of the SU(2) simulations. At 11.7$A$\,GeV the hyperon spectra
are closer to the SU(3) calculations and for $\Omega^-+\bar\Omega^+$ lie even below
those.

Another interesting observable measured in experiment is the
transverse mass ($m_t$) spectrum at midrapidity, i.e. $dN/(m_t dy
dm_t)$ as displayed in Fig. \ref{fig:mtspectrum}.
Here the additional strangeness sector has qualitatively the same
impact as for the rapidity spectra. Accordingly, we only show
results for central Pb+Pb  collisions in the energy regime from 20
to 158$A$\,GeV including the strangeness sector for the $B\bar
B\leftrightarrow 3M$ reactions. For the  $\Xi^-$ we find that PHSD
describes the low $m_t$ regime for energies below 158$A$\,GeV rather
well. However, for higher $m_t$ the data points are missed due to a
harder experimental slope of the spectrum. At 158$A$\,GeV some
$\Xi^-$'s are missed in the low $m_t$ regime. The $\bar\Lambda+\bar\Sigma^0$
spectrum is close to the experimentally measured data for all energies,
however, at 158$A$\,GeV it falls off too fast. The transverse mass
spectra of the antiprotons are overall in very good agreement with
experiment, the only drawback is the overproduction at midrapidity
which is most visible for 20 and 30$A$\,GeV. Also, the $\bar\Xi^+$
are in close vicinity to the experimental data for energies smaller
than 158$A$\,GeV, but fall off too quickly at 158$A$\,GeV. The
production of $\Omega^-$ and $\bar\Omega^+$ was underestimated
already in the rapidity spectra, see Fig. \ref{fig:energyrange2},
but looking at the transverse mass spectra at
158$A$\,GeV the results are in reasonable agreement with experiment
for  $m_t<0.8$ GeV.

\subsection{Impact of chiral symmetry restoration and deconfinement}
\begin{figure*}[t]
\centering
\includegraphics[width= 0.43\textwidth]{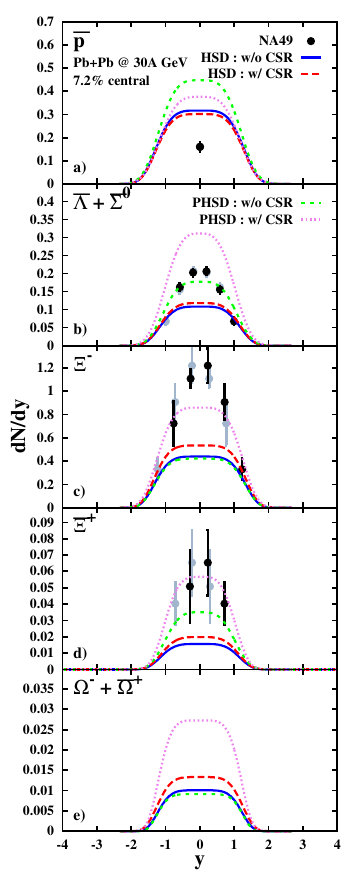}\hspace{0.5cm}
\includegraphics[width= 0.43\textwidth]{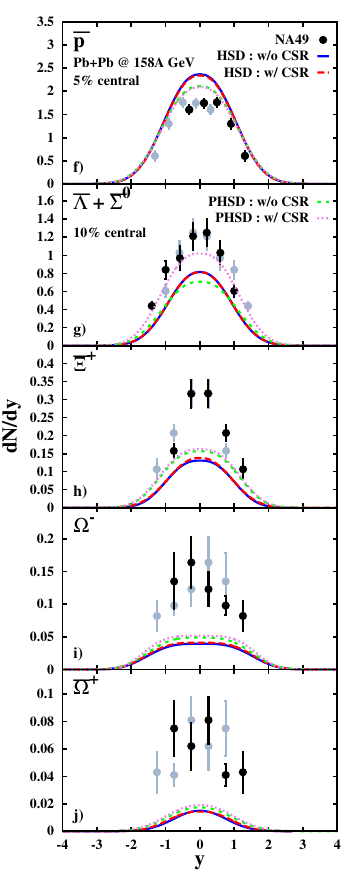}
\caption{(Color online) Rapidity spectra for a central Pb+Pb collision at 30 and
158$A$\,GeV; comparison between simulations with (PHSD) and without
(HSD) the deconfinement transition and with activated and
deactivated chiral symmetry restoration (CSR). The data points are
taken from Refs. \cite{Anticic:2010mp,Alt:2008qm,Alt:2004kq,Alt:2006dk}.}\label{fig:CSR}
\end{figure*}
\begin{figure}[t]
{\centering
\includegraphics[width= 0.35\textwidth]{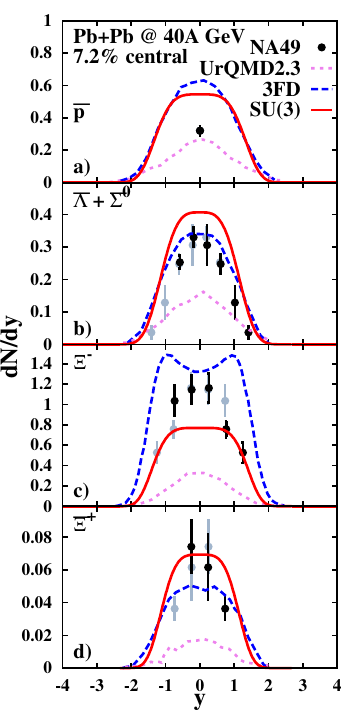}}
\caption{(Color online) Rapidity spectra for a central Pb+Pb collision at
40$A$\,GeV; comparison between PHSD results with the $B\bar
B\leftrightarrow3M$ reactions including strangeness (red solid
line), UrQMD-2.3 \cite{Petersen:2008kb} (violet short-dashed line) and 3FD
with a 2-phase equation of state \cite{Ivanov:2013yqa} (blue dashed line). The
experimental data are taken from Refs.
\cite{Anticic:2010mp,Alt:2008qm}.
}\label{fig:comparisonPHSDothermodels}
\end{figure}
We now address the question with respect to traces of chiral
symmetry restoration and deconfinement in the antibaryon and
multi-strange baryon spectra from central Pb+Pb collisions at SPS
energies. We recall that clear signals have been found before in the
strange meson and baryon rapidity distributions
\cite{Cass16,AlesPaper} and one might speculate if a similar signal
can be seen in the antibaryon sector. To this aim we perform
transport calculations - including the $B {\bar B} \leftrightarrow
3M$ channels specified above - with different settings:
\begin{itemize}
\item{HSD calculations without chiral symmetry restoration (CSR)
and deconfinement since HSD does not include a partonic phase }
\item{HSD calculations with chiral symmetry restoration (CSR) in the
hadronic phase but without deconfinement}
\item{PHSD calculations without chiral symmetry restoration (CSR) in the
hadronic phase but with a deconfinement transition}
\item{PHSD calculations with chiral symmetry restoration (CSR) in the
hadronic phase and with a deconfinement transition}.
\end{itemize}
The systems addressed are central Pb+Pb collisions at 30 and
158$A$\,GeV. The rapidity spectra for antibaryons and $\Xi^-$ are
displayed in Fig. \ref{fig:CSR} and show that at 158$A$\,GeV the
impact of chiral symmetry restoration is very small in the HSD
calculations (without deconfinement) as well as for PHSD (including
deconfinement) except for the ${\bar \Lambda}+{\bar \Sigma}^0$
spectrum. When comparing HSD and PHSD results including CSR we find
a slight reduction of the ${\bar p}$ spectra, a moderate enhancement
for the ${\bar \Lambda}+{\bar \Sigma}^0$ spectrum and only a small
enhancement for $\Xi^\pm$ and $\Omega^- + {\bar \Omega}^+$ when
including a partonic phase. Since the reproduction of the
multistrange sector by PHSD is very poor one cannot conclude on the
presence of a deconfinement transition on the basis of the rapidity
spectra shown in Fig. \ref{fig:CSR}. Note, however, that a clear
signal has been found in the elliptic and triangular flow before in
Ref. \cite{Konchakovski:2012yg} at this energy.

At 30$A$\,GeV the situation is not much better. The PHSD
calculations with CSR perform best for $\Xi^-$ and ${\bar \Xi}^+$,
however, overestimate the ${\bar p}$ and ${\bar \Lambda}+{\bar
\Sigma}^0$ yield. The HSD calculations are too low in the strange
antibaryon sector including/excluding CSR providing some hint that a
partonic phase should be present in a moderate space-time volume at
this energy. Accordingly, the antibaryons and in particular the
multi-strange sector do not give additional information on chiral
symmetry restoration or deconfinement within the framework of PHSD
calculations.

\subsection{Comparison to other dynamical models}
In this subsection we compare our current PHSD results to those from
other dynamical models which have been employed for heavy-ion
reactions in the SPS energy regime, in particular from the
Ultra-relativistic Quantum Molecular Dynamics model (UrQMD)
\cite{Bass:1998ca,Bleicher:1999xi} and the three-fluid dynamics
model (3FD) \cite{3FD}. The UrQMD is a hadronic transport model
including a multitude of hadronic resonances as well as strings that
are responsible for multi-particle production. The 3FD is a fluid
dynamical model describing - within the framework of hydrodynamics -
the transition from the initial baryonic fluids (projectile and
target) to the newly produced fluid (around midrapidity). For
details we refer the reader to the original literature
\cite{Bass:1998ca,Bleicher:1999xi,3FD}.  We show in Fig.
\ref{fig:comparisonPHSDothermodels} our actual results in case of
the rapidity spectra for a central Pb+Pb collision at 40$A$\,GeV
with the $B\bar B\leftrightarrow 3M$ reactions including the
strangeness sector in comparison to results from the UrQMD
\cite{Petersen:2008kb} and the 3FD using a 2-phase equation of state
\cite{Ivanov:2013yqa}. The 3FD model, like PHSD, overshoots the
antiproton yield whereas UrQMD is close to the experimental data.
The $\bar \Lambda+\bar \Sigma^0$ spectrum is described by PHSD and
the 3FD model similarly close to the experimental data whereas UrQMD
produces too few. For the $\Xi^-$ all models show different
behaviors; whereas the 3FD model overpredicts the production, PHSD
produces slightly too few $\Xi^-$ at midrapidity but describes
otherwise the shape well. UrQMD predicts (just like for
$\bar\Lambda+\bar\Sigma^0$ and $\bar\Xi^+$) too few antibaryons
since $B{\bar B}$ annihilation is incorporated, however, not the
backward channels thus violating detailed balance. PHSD and the 3FD
model are close to the experimental data for $\bar\Xi^+$, with the
3FD slightly underpredicting the yield. Depending on the particle
species of interest one model describes some yield better than the
other at higher SPS energies. In general, the 3FD model and PHSD
appear to be similarly capable of roughly describing the dynamics of
baryons and antibaryons with strangeness content in this energy
range.

\section{Summary}\label{sec:summary}

In this work we have recapitulated and extended the quark
rearrangement model for baryon-antibaryon annihilation ($B {\bar B}
\leftrightarrow 3 M$) in the course of heavy-ion collisions. The
approximate validity of this model was motivated by the distribution
in the number of final state pions in $p\bar p$ annihilation for
$2.3$\,GeV$\leq\sqrt s\leq 4$\,GeV (cf. Fig. 1), where the 3-body
channel $\pi \rho \rho$ e.g. leads to 5 pions (on average) in the
final state. Additionally to the HSD calculations in Ref. \cite{Cassing:2001ds}, we have
included in the $2 \leftrightarrow 3$ channels the strangeness sector with a suppression factor for the
matrix elements of particles having strange and anti-strange quarks.
We have shown, using simulations in a box with periodic boundary
conditions, that the numerical implementation of the quark
rearrangement model including the strangeness sector satisfies the
detailed balance $2 \leftrightarrow 3$ relation on a
channel-by-channel basis as well as differentially as a function of
the invariant energy $\sqrt{s}$.

We found that the earlier rates from HSD2.3 \cite{Cassing:2001ds} differ substantially
from the present results from PHSD (version 4.0) due to the different degrees of freedom in
the initial phase of the collision.  Both rates (from HSD2.3 and PHSD4.0) differ only slightly
for times $\geq$ 6 fm/c (after contact of Pb+Pb at b=2fm) but the huge rates (from HSD2.3) at the first
few fm/c are essentially missing in PHSD4.0. This is due to the fact
that at the top SPS energy the initial energy conversion goes to interacting partons
in PHSD4.0 and not to strings decaying to hadrons (and partly to $B {\bar B}$ pairs) in HSD2.3.
Thus in PHSD4.0 (at the top SPS and higher energies) there are initially no $B {\bar B}$ pairs that
might annihilate nor mesons that might fuse! Due to the very high hadron densities in HSD2.3
(after string decay) both the annihilation and reproduction rates are very high and about
equal whereas in the hadronic expansion phase the densities are sizeably lower. In this
dilute regime the 3-body channels first dominate and decrease fast in time whereas the 2-body
annihilation reactions still continue for some time.
However, in both transport calculations -- incorporating the $2 \leftrightarrow 3$ reactions --
the time integrated rates for annihilation and reproduction turn out to be about equal.

The influence of the newly implemented channels in the strangeness
sector on actual heavy-ion collisions has been investigated in PHSD
simulations (version 4.0) of central Pb+Pb collisions from 11.7-158$A$\,GeV. The
rapidity spectra of antibaryons - using the quark rearrangement
model with and without the strangeness sector - have been compared
to experimental data where available. Changes could only be seen for
the antibaryons in the investigated energy regime whereas the meson
and baryon sector are practically unchanged \cite{AlesPaper}. Due to
the chemical rearrangement between the baryons and mesons considered
in $B\bar B\leftrightarrow 3M$ reactions an overall higher
anti-proton production  was observed for all energies which pushed
the PHSD results up thus overestimating the experimental data. The
other antibaryons got closer to the experimental data when the
strangeness sector was included for the $B\bar B\leftrightarrow 3M$
reactions. For the  energies investigated the strangeness sector has
the largest impact at the lowest energies. The results show that the
$B\bar B\leftrightarrow 3M$ reactions indeed need the strangeness
sector to describe the heavy-ion collisions more properly. We note,
however, that the quark rearrangement model might be too crude to
allow for robust conclusions. We still need experimental information
on baryon-antibaryon annihilation cross sections other than $p\bar
p$ and $p\bar n$ to achieve a better description and understanding of heavy-ion
collisions.

In addition to the rapidity spectra, we have shown the transverse
mass spectra for various antibaryons and have seen that the low
$m_t$ region is well described for all antibaryons with the
exception of the antiprotons that are overpredicted at energies
lower than 80$A$\,GeV. For higher transverse masses some spectra
fall off too fast thus underestimating  the experimental data to
some extent. Accordingly, our understanding of antibaryon dynamics
is far from being complete and we might still miss essential
ingredients.

We have additionally addressed the question if the antibaryon
spectra (with strangeness) from central heavy-ion reactions at SPS
energies provide further information on the issue of chiral symmetry
restoration and deconfinement. By comparing results from HSD
(without partonic phase) with those from PHSD (with  partonic
degrees of freedom) as well as including/excluding effects from
chiral symmetry restoration (Fig. \ref{fig:CSR}) we did not find
convincing signals for either transition  due to the strong
final-state interactions.

\begin{acknowledgments}
The authors acknowledge inspiring discussions with E. L. Bratkovskaya,
P. Moreau, A. Palmese and T. Steinert. We
thank the Helmholtz International Center for FAIR (HIC for FAIR),
the Helmholtz Graduate School for Hadron and Ion Research
(HGS-HIRe), the Helmholtz Research School for Quark Matter Studies
in Heavy-Ion Collisions (H-QM) for support. The computational
resources have been provided by the Center for Scientific Computing
(CSC) in the framework of the Landes-Offensive zur Entwicklung
Wissenschaftlich-\"okonomischer Exzellenz (LOEWE) and the Green IT
Cube at FAIR.
\end{acknowledgments}

\appendix
\section{Meson fusion}\label{sec:mesonfusion}
In order to determine the probability for the three-meson fusion
rate we start with the Lorentz-invariant reaction rate for this
process \cite{Cassing:2001ds},
\begin{align}
\begin{split}
&\frac{\mathrm{d}N_\mathrm{coll}[3~\mathrm{mesons} \rightarrow B\bar B]}{\mathrm{d}t\,\mathrm{d}V}=\\
&\sum_c\sum_{c'}\frac{1}{(2\pi)^9}\int\frac{\mathrm{d}^3p_3}{2E_3}\frac{\mathrm{d}^3p_4}{2E_4}\frac{\mathrm{d}^3p_5}{2E_5}~W_{2,3}(\sqrt{s})\\ &\times   R_2(p_3+p_4+p_5;c') N_B^{c'} f_3(x,p_3)f_4(x,p_4)f_5(x,p_5),\end{split}\label{eq:lorentzinvariant32}
\end{align}
where  $N_B^{c'}$ denotes the multiplicity of the final state and
the two-body phase-space integral  $R_2$ is given by
\begin{align}
R_2(\sqrt s;m_1,m_2)=\frac{\sqrt{\lambda(s,m_1^2,m_2^2)}}{8\pi s}
\end{align}
with $\lambda$ defined in Eq. (\ref{eq:lambdadef}). The transition
matrix element squared $W_{2,3}$ is not known but using Eq.
(\ref{eq:crosssection}) for our special problem of $2\leftrightarrow
3$ processes one gets,
\begin{align}
\begin{split}
\sum_c P_{c\rightarrow c'}(\sqrt s)=&\sum_c W_{2,3}(\sqrt s) R_3(\sqrt s,c)N_\mathrm{fin}^c \\= &W_{2,3}N_3^{-1}(\sqrt s,c')\\=&4E_1E_2 v_\mathrm{rel}\sigma_\mathrm{ann}^{c'}(\sqrt s),
\end{split} \label{eq:MatrixelementAndCrosssection}
\end{align}
where we have taken $W_{2,3}$ out of the sum over the
baryon-antibaryon pairs and end up with the expression for the
normalisation constant for the invariant energy $\sqrt s$ from Eq.
(\ref{eq:N3}). Inserting Eq. (\ref{eq:MatrixelementAndCrosssection})
for the transition matrix element squared into
(\ref{eq:lorentzinvariant32}) gives the result for the transition
probability for the meson fusion in Eq. (\ref{eq:P32}). Note that
all energies and momenta in the calculations of the transition
probabilities are in the lab frame.
\section{Phase-space integrals}\label{sec:phasespace}
The on-shell phase-space integrals occurring throughout this work
inhibit most of the dynamics of the system. As they play a major
role this section is dedicated to some more details of phase-space
integrals. We recall that the $n$-body phase-space integral is
generally defined by
\begin{multline}
 R_n(P;m_1,\ldots,m_n)=\\
 \left(\frac{1}{(2\pi)^3}\right)^n \int\prod_{k=1}^n \mathrm{d}^4p_k~\rho_k(p_k) (2\pi)^4\delta^4\left(P-\sum_{j=1}^n p_j\right),
\label{eq:phasespace-general}
\end{multline}
with $\rho$ denoting the spectral function of the respective
particle. Since the phase-space integrals are Lorentz invariant we
will always work in the center-of-mass system. In the on-shell case
the spectral function takes the form
\begin{equation} \label{onsh} \rho(p)=\delta(p^2-m_1^2) \end{equation}
with $p$ denoting the 4-momentum in this case. Inserting the
spectral function (\ref{onsh}) into Eq.
(\ref{eq:phasespace-general}) and integrating over  $p^0$  yields
the on-shell phase-space integral of Eq. (\ref{eq:n-phasespace}). To
show (as an example) the behavior of the different $n$-body
phase-space integrals it is instructive to look e.g. at the
consecutive decays $p\bar p\rightarrow \pi\rho\rho \rightarrow
3\pi\rho \rightarrow 5\pi$ which are essentially the motivation for
the QRM. Also, this example connects  the 3-,4- and 5-body
phase-space integrals as a function of the invariant energy above
threshold (see below).

\begin{figure}[t]
\centering
\includegraphics[width= 0.48\textwidth]{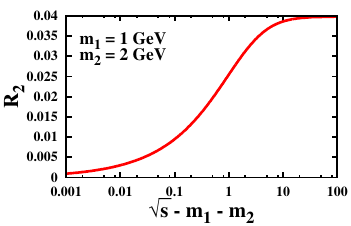}
\caption{Two-body phase-space integral for particles with masses
$m_1=1$\,GeV and $m_2=2$\,GeV as a function of the invariant energy
above threshold.}\label{fig:R2}
\end{figure}
\begin{figure}[b]
{\centering
\includegraphics[width= 0.48\textwidth]{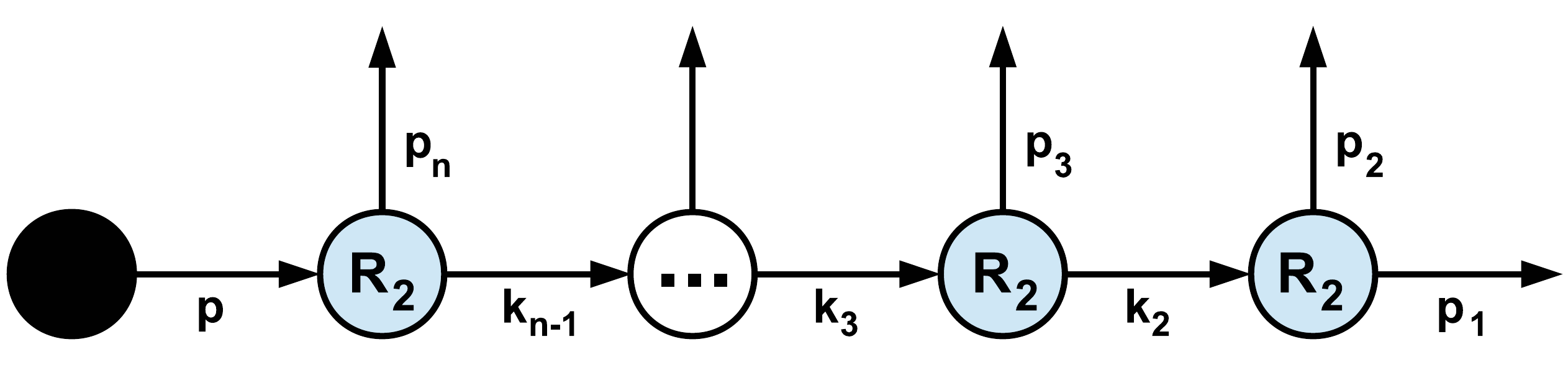}}
\caption{Illustration of the subsequent decay of an initial state
(black dot) into $n$ particles. The initial state may consist of $m$
particles as only the invariant mass is relevant for the phase-space
integral due to  Lorentz invariance.}\label{fig:subsequentDecay}
\end{figure}
\begin{figure}[t]
{\centering
\includegraphics[width= 0.48\textwidth]{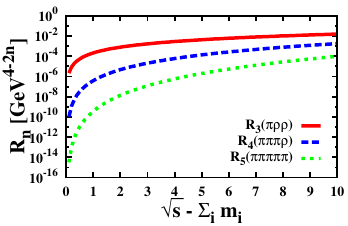}}
\caption{(Color online) Illustration of the 3-, 4-, and 5-body
phase-space integrals as a function of the invariant energy above
threshold. The red solid line shows the 3-body phase-space integral
for $\pi\rho\rho$, the blue dashed line shows the 4-body phase-space
integral for $3\pi\rho$ and the green dashed line shows the 5-body
phase-space integral for 5 pions.}\label{fig:R345}
\end{figure}
\begin{figure}[b]
{\centering
\includegraphics[width= 0.48\textwidth]{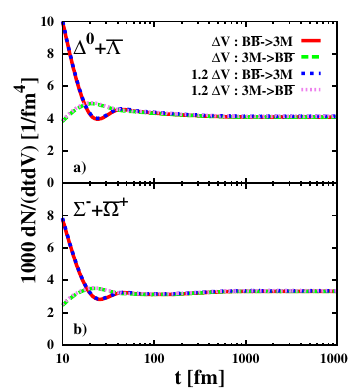}}
\caption{(Color online) Concistency check for a change in the cell
size $\Delta V$ by 20\%. a) for $\Delta^0+\bar\Lambda$ and b) for
$\Sigma^-+\bar\Omega^+$ initalizations. The red solid line shows the
baryon-antibaryon annihilation for the cell volume $\Delta V$, the
green dashed line shows the baryon-antibaryon formation for $\Delta
V$, the blue short-dashed line shows the baryon-antibaryon
annihilation for $1.2\Delta V$ and the violet dotted line shows the
baryon-antibaryon formation for $1.2\Delta V$.}\label{fig:cellsize}
\end{figure}
\begin{figure}[b]
{\centering
\includegraphics[width=
0.48\textwidth]{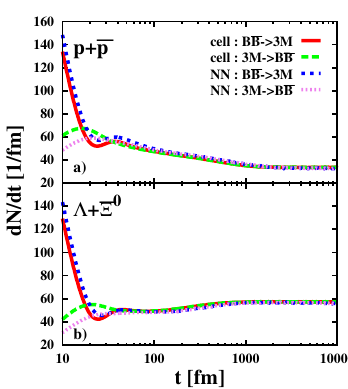}}
\caption{(Color online) Comparison of the reaction rates between the cell algorithm (cell) and the
next-neighbor (NN) realization of the in-cell method.  The systems shown are in a) the $p+\bar p$ and in b) the
$\Lambda+\bar\Xi^0$ initialization. The red solid line shows the
baryon-antibaryon annihilation for the cell method, the green dashed
line shows the baryon-antibaryon formation for the cell method, the
blue short-dashed line shows the baryon-antibaryon annihilation for
the NN method and the violet dotted line shows the baryon-antibaryon
formation for the NN method.}\label{fig:cell-NN}
\end{figure}

For the sake of completeness, we start with the 1-body phase-space
integral,
\begin{align}
R_1(\sqrt s;m)=\frac{1}{(2\pi)^3}\int \frac{\mathrm{d}^3p}{2E}~(2\pi)^4\delta^4(\sqrt s - E) = \frac{\pi}{\sqrt s},
\end{align}
where E is the on-shell energy $E=\sqrt{m^2+p^2}$ and the mass $m$
of the particle is equal to the invariant energy $\sqrt s$. This
result shows that the 1-body phase-space decreases with increasing
$\sqrt s$. The 2-body phase space can also be evaluated
analytically,
\begin{align}
\begin{split}
R_2&(\sqrt s; m_1,m_2)=\\
 &~\frac{1}{(2\pi)^2}\int\int \frac{\mathrm{d}^3p_1}{2E_1}\frac{\mathrm{d}^3p_2}{2E_2}~\delta^3(\vec p_1+\vec p_2)\delta(\sqrt s -E_1-E_2)\end{split}\\
=& \frac{1}{4(2\pi)^2} \int \frac{\mathrm{d}^3p_1}{E_1E_2}\delta(\sqrt s -E_1-E_2)\\
=& \frac{1}{4(2\pi)^2} \int\limits_{0}^{\infty}\int\limits_0^\pi\int\limits_0^{2\pi} \frac{\mathrm{d}\phi\mathrm{d}\theta\mathrm{d}p_1~p_1^2\sin\theta}{E_1E_2}  \delta(\sqrt s -E_1-E_2)\\
=& \frac{1}{4\pi} \int\limits_{0}^{\infty} \frac{\mathrm{d}p_1~
p_1^2}{\sqrt{m_1^2+p_1^2}\sqrt{m_2^2+p_1^2}} \delta\left(\sqrt s
-E_1-E_2\right).\label{eq:R2-middle}
\end{align}
The zeroes of the delta function are given by
\begin{align}
p_0=\pm \frac{\sqrt{\lambda(s,m_1^2,m_2^2)}}{2\sqrt s},\label{eq:R2-p0}
\end{align}
where only the positive value has to be taken in our calculation.
Rewriting the delta function as
\begin{align}
\delta(\sqrt s -E_1-E_2)=\frac{\delta(p_1-p_0)}{p_1/E_1+p_1/E_2}
\label{eq:R2-delta}
\end{align}
and plugging Eqs. (\ref{eq:R2-p0}) and (\ref{eq:R2-delta}) into Eq.
(\ref{eq:R2-middle}) we obtain the two-body phase-space integral
\begin{align}\begin{split}
R_2(\sqrt s; m_1,m_2)=&\frac{1}{4\pi} \int\limits_{0}^{\infty} \frac{\mathrm{d}p_1~ p_1}{E_1E_2}  \frac{E_1E_2\delta(p_1-p_0)}{E_1+E_2}\\
=&\frac{\sqrt{\lambda(s,m_1^2,m_2^2)}}{8\pi s},\label{eq:R2}
\end{split}\end{align}
with $E_1+E_2=\sqrt s$ from the original delta function. The typical
shape of $R_2(\sqrt{s},m_1,m_2)$ is shown in Fig. \ref{fig:R2} for
the masses $m_1=1$\,GeV and $m_2=2$\,GeV as a function of the
invariant energy above threshold. The upper limit is independent of
the masses and is given by $1/8\pi$.

The on-shell three-body phase-space integral $R_3(\sqrt{s},
m_1,m_2,m_3)$ is the most important one for our work and a good
example for the evaluation of  phase-space integrals of higher order
since the $n$-body decay can be considered as consecutive 2-body
decays, see Fig. \ref{fig:subsequentDecay} for an illustration. Note
that in Fig. \ref{fig:subsequentDecay} $k_n=p$ and $k_1=p_1$. A
prerequisite in calculating the phase-space integral  is that we do
not have any incoming momenta in between the first and final 2-body
decay. For the calculation of the process we employ the recursion
relation for phase-space integrals,
\begin{align}
R_n(P)=\int\frac{\mathrm{d}^4p_n}{(2\pi)^3}\rho_n(p_n)R_{n-1}(P-p_n),\label{eq:Rn-recursive}
\end{align}
and also  insert two identities
\begin{align}
1=&\int \mathrm d M_{n-1}^2 \delta(M_{n-1}^2-k_{n-1}^2)),\label{eq:one1} \\
1=&\int\mathrm d^4 k_{n-1} \delta^4(P-p_n-k_{n-1}).\label{eq:one2}
\end{align}
The first identity from Eq. (\ref{eq:one1}) gives the mass of the
first cluster from which the 4-momentum $p_n$ splits. The second
identity ensures energy-momentum conservation in the splitting
process. Plugging both identities into Eq. (\ref{eq:Rn-recursive})
we find
\begin{widetext}
\begin{align}
R_n(P)=&\int\mathrm{d}M^2_{n-1}\underbrace{\int\mathrm{d}^4k_{n-1}\int\frac{\mathrm{d}^4p_n}{(2\pi)^3} \delta^4(k_{n-1}^2-M_{n-1}^2) \delta^4(p_n^2-m_n^2)\delta^4(P-p_n-k_{n-1})}_{R_2(P;m_n,M_{n-1})/(2\pi)} R_{n-1}(k_{n-1})\\
=&\int\limits_{(\sum_{i=1}^{n-1}m_i)^2}^{(M_n-m_n)^2} \mathrm{d}M^2_{n-1} \frac{R_2(P;m_n,M_{n-1})}{2\pi} R_{n-1}(k_{n-1}).\label{R2-recursive}
\end{align}\end{widetext}

\begin{figure}[b]
{\centering
\includegraphics[width= 0.48\textwidth]{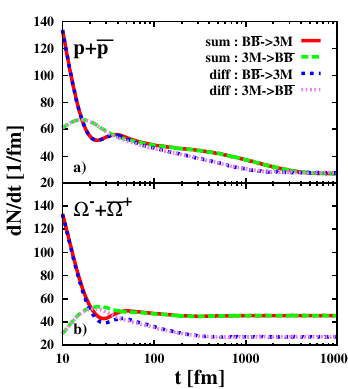}}
\caption{(Color online) Comparison of the reaction rate between the sum and the
difference of the strange and antistrange quarks in the calculation
of transition probabilities in $B \bar B\leftrightarrow 3M$
reactions (denoted by sum and diff). a) shows the $p+\bar p$ and b)
the $\Omega^-+\bar \Omega^+$ initialization. The red solid line
shows the baryon-antibaryon annihilation of the sum, the green
dashed line shows the baryon-antibaryon formation of the sum, the
blue short-dashed line shows the baryon-antibaryon annihilation of
the difference and the violet dotted line shows the
baryon-antibaryon formation of the
difference.}\label{fig:strangeness}
\end{figure}
\begin{figure}[h!]
\centering
\includegraphics[width= 0.35\textwidth]{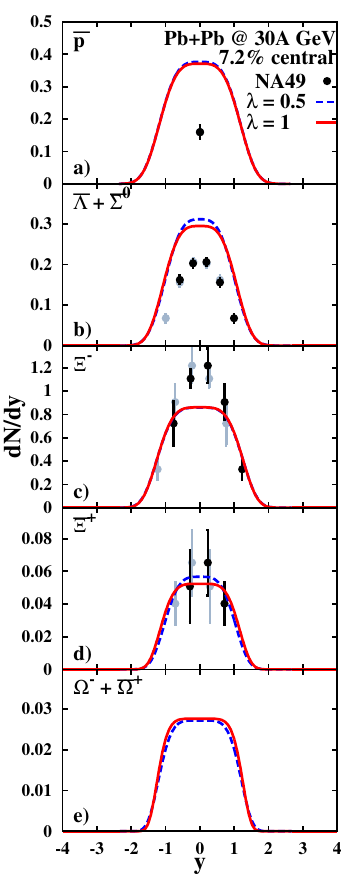}
\caption{(Color online) Rapidity spectra for a central Pb+Pb
collision at 30$A$\,GeV; comparison between simulations with a
strangeness suppression factor $\lambda=0.5$ (dashed lines) and no
strangeness suppression, i.e.  $\lambda=1$ (solid
lines).}\label{fig:suppressionfactors}
\end{figure}

With this expression any $n$-particle phase-space integral can be
calculated in a straight forward fashion as long as the masses $m_i$
are known.  Note that the last $R_2$, which one gets after applying
Eq. (\ref{R2-recursive}) several times, has no additional factor
$1/(2\pi)$. In Fig. \ref{fig:R345} the phase-space integrals  for 3,
4 and 5 particles are shown as a function of the invariant energy
above threshold for our example of initial $\pi\rho\rho$ with a
subsequent decay into $3\pi\rho$ and a final decay to 5 pions. All
phase-space integrals share a similar shape, only the magnitudes
close to threshold vary substantially with the number of particles.

\section{In-cell method: cell-size dependence}\label{sec:cellsize}
We here show the stability of our approach with respect to the
equilibrium state when changing the size of the cells. For this
investigation we keep the time step d$t$ constant but enhance the
cell volume $\Delta V$ by 20\% and compare the reaction rate as a
function of time to the default calculations in Fig.
\ref{fig:cellsize}. We observe that the change in the cell size does
not have any impact on the equilibration at all. For all times  both
cell sizes produce the same results giving testimony to the
stability of the numerical implementation.

\section{In-cell method versus next-neighbor interaction}\label{sec:incell-NN}
The in-cell method used for the description of the $B\bar
B\leftrightarrow 3M$ reactions has been implemented  cutting
effectively the space-time into cells of cell-size $\Delta V \times
\Delta t$ and letting only particles of the same cell interact with
each other. Another possibility for the implementation of the
baryon-antibaryon annihilation (and recreation) is by defining the
volume $\Delta V$ by a sphere around the first particle and letting
all particles in the sphere interact with each other; this
implementation we denote by next-neighbor (NN) algorithm in the
following. In Fig. \ref{fig:cell-NN} we compare the results of these
two choices. Due to the large finite size effects for the NN method
the volume of the box had to be enhanced and filled with the same
density as the standard box but  letting only the particles inside
the standard box volume be the particles from whose sphere
the partners are selected. After employing this
minimization of finite size effects we find that both methods give
the same reaction rates for times larger than $\approx$30\,fm. A
small deviation between both methods is seen for smaller times. As
expected one might use in general also the NN method. The
disadvantage of the numerical implementation of the NN method is the
larger computational time in comparison to the discretization of
space-time.  Thus PHSD uses the in-cell method not for the
individual cells from the NN method but for the fixed cells of the
space-time discretization.

\section{Strangeness suppression}\label{sec:strangeness}
A further point to discuss in our model is whether to use the sum or
the difference of the number of strange and anti-strange quarks in
Eq. (\ref{eq:crosssection}) for the strangeness suppression. Fig.
\ref{fig:strangeness} illustrates the deviation between the two
suppression models  for the total reaction rate. For the system
consisting initially only of light quarks, $p+\bar p$, we see no
sizeable differences between the sum and the difference of strange
and anti-strange quarks in Eq. (\ref{eq:crosssection}). The system
with an initial large difference between the number of strange and
anti-strange quarks, $\Omega^-+\bar\Omega^+$, converges to rather
different equilibrium states for the two assumptions. The
suppression with the sum leads to an overall larger total reaction
rate and its equilibrium value is twice as large as the suppression
with the difference assumption. However, both models produce rather
similar results for times $t<50$\,fm, which is of relevance for the
heavy-ion collisions considered in this work. Accordingly  we use in
PHSD the suppression with the sum of the number of strange and
anti-strange quarks since both models give practically identical
results in PHSD simulations of relativistic heavy-ion reactions.

Another issue relates to the actual value of the strangeness suppression factor $\lambda$
which had been taken as $\lambda$ = 0.5. In order to demonstrate the impact
of the parameter $\lambda$ on antibaryon spectra we show in Fig. \ref{fig:suppressionfactors} the
rapidity distributions for central Pb+Pb collisions at 30 A GeV for $\lambda$=0.5
(dashed lines) and $\lambda$=1 (solid lines).
Without strangeness suppression in $2\leftrightarrow 3$ reactions for hadrons with strange/antistrange quarks we find at
30$A$\,GeV that the rapidity spectra of $\bar\Lambda+\bar\Sigma^0$ and $\bar\Xi^+$ are slightly shifted to lower values
and broadened in comparison to the standard value of $\lambda=0.5$. The spectrum for $\Omega^-+\bar \Omega^+$
very slightly broadens and the $\bar p$ spectrum is basically not influenced by the change of $\lambda$.

%\bibliography{references}{}
%\bibliographystyle{apsrev4-1}

\end{document}